\documentclass[%
 reprint,
 amsmath,amssymb,
 aps,
pra,
floatfix,
showkeys
]{revtex4-2}
\usepackage[english]{babel}

\usepackage{graphicx} 

\usepackage{amsthm} % for proofs
\newtheorem{theorem}{Theorem}[section]   % for theorems

\usepackage{qcircuit, braket}
\usepackage[ruled,vlined]{algorithm2e}

\usepackage{graphicx}
\usepackage{subcaption} % for subfigures
\usepackage[margin=1in]{geometry}

\begin{document}

\preprint{APS/123-QED}

\title{Universal Blind Quantum Computation with Recursive Rotation Gates}

\author{Mohit Joshi}
    \email{joshimohit@bhu.ac.in}
\author{Manoj Kumar Mishra}%
 \email{mkmbhuvns@bhu.ac.in}
\author{S. Karthikeyan}%
 \email{karthik@bhu.ac.in}
\affiliation{%
Department of Computer Science, \\
Banaras Hindu University, Varanasi, India - 221005
}%

\begin{abstract}
Blind Quantum Computation lets a limited-capability client delegate its complex computation to a remote server without revealing its data or computation. Several such protocols have been proposed under varied quantum computing models. However, these protocols either rely on highly entangled resource states (in measurement-based models) or are based on non-parametric resource sets (in circuit-based models). These restrictions hinder the practical applicability of such an algorithm in the NISQ era, especially concerning the hybrid quantum-classical infrastructure, which depends on parametric gates. We present a protocol for universal blind quantum computation based on recursive decryption of parametric rotation gates, which does not require a highly entangled state at the server side and substantially reduces the communication rounds required for practical prototyping of secure variational algorithms.
\end{abstract}

\maketitle

\section{Introduction}
With the anticipated interference of quantum-enabled internet, the applications of secure networking will broaden its horizon beyond classical limits \cite{wehner_quantum_2018}. This enables the opportunity for fundamentally secure and private internet based on the information-theoretic principles \cite{fitzsimons_private_2017, joshi2022recent}.
These advances are evident from the availability of quantum service providers like IBM, D-Wave, Rigetti, and Microsoft, among many others \cite{moguel_quantum_2022}. 

One such prospect of security is the secure delegation of quantum computation to a remote server. This paradigm enables a limited-capability client to perform complex quantum computation securely. This enables the foundation of an unconditionally secure distributed quantum internet.
Such a paradigm of security for data can be provided using Pauli's $X$ and $Z$ rotation gates assisted by classical random bits \cite{childs_secure_2005}. These decryption of gates in such protocols utilised the commutation properties of $X$ and $Z$ with respect to the commuting gate. The gate from Clifford set can be implemented without interaction on the server \cite{arrighi_blind_2006}, while non-Clifford gates require additional interaction between client and server \cite{broadbent_delegating_2015,tan_universal_2017}.

The security of computation in such protocols is provided by performing computation over some universal resource set using Childs' encrypted data. Broadbent et al. first proposed such a resource set, called brickwork state, in a measurement-based quantum computing model \cite{broadbent2009universal}. Since then, many variants of measurement-based resource sets have been proposed \cite{morimae_blind_2012,morimae_ground_2011}.

However, such models require a highly entangled resource state at the server side to perform meaningful computation. Recent years have seen several optimisations to the measurement-based resource set \cite{ma_universal_2024,zhang_hybrid_2019}. Models based on ancilla-driven approaches have also been proposed \cite{sueki2013ancilla}. 

Circuit-based models require the least amount of server overhead, and several protocols for circuit-based universal resource sets have been proposed \cite{zhang_single-server_2018,liu_full-blind_2020,sano_blind_2021,zhang_gate_2022}. These protocols have started the implementation of blind circuits in various fields \cite{liang2015quantum, qu2022secure, liu2019quantum, joshi2024leveraging}.

However, these approaches are based on the decryption of non-parametric resources. This requires the parametric circuit to be decomposed to the gate set $\{H,S,T,CX,CZ,CCX\}$ before decryption. The decomposition of parametric gates inadvertently increases the overall depth of the circuit to be implemented for blind delegation of the algorithm. Moreover, in the noisy-intermediate scale quantum (NISQ) era, parametric gates can be natively implemented in the hardware \cite{mckay2017efficient, chen_compiling_2023,javadi2024quantum}.

Hence, the direct decryption of a parametric gate without prior decomposition can massively reduce the resources required at the server and the communication cost of such protocols. In recent work \cite{joshi2025quantum}, such a technique of recursive decryption of $R_z$ gates has been proposed. However, this approach provides blindness to only data. In this study, we extend this protocol to universal blindness of data and computation by hiding the outgoing classical information shared between client and server. The main contributions of this study are:
\begin{itemize}
    \item[1] We proposed the procedure for blind decryption of $R_z(\theta)$ using $\upsilon=(p_o\pi + \pi - \pi/2^M)$ where $M=\lceil \log_2(\pi/\epsilon)\rceil$ which depends on the precision of computation $\epsilon$ and independent of algorithm running on the server.
    \item[2] We proposed a novel resource state $J = (H_1,CZ_{2,3},R_z(v))$ for universal blind quantum computation by hiding $\theta$ utilized in the computation. 
    \item[3] We show this protocol requires atmost $\mathcal{O}((n_p + n_{np})\log_2^2(\pi/\epsilon))$ communication rounds while the protocols based on non-parametric resources requires $\mathcal{O}(n_{p}\ln^{3.97}(1/\epsilon) + n_{np})$ rounds, where $n_p:n_{np}$ is the ratio of parametric and non-parametric gates in a given algorithm.
\end{itemize}

The rest of the study is organised as follows: 
In section \ref{sec:background}, we present the background necessary to understand the subject. In section \ref{sec:proposed} we present the proposed algorithm. In section \ref{sec:compartive_analysis} we present a comparative analysis of our algorithm with other circuit-based algorithms.
At last \ref{sec:conclusion} presents the concluding remarks. 

\section{Background}\label{sec:background}
Child proposed the idea that quantum information encrypted using the randomly chosen element from the set $\{I,X,Z,XZ\}$ is a quantum analogue of a one-time pad, as the trace of the density matrix is maximally mixed:
\begin{align}
2^n \mathbb{I} = \sum_{j_1, j_2, ..., j_{2n} \in\{ 0,1\}} \bigg(\bigotimes_{i=1}^n Z_i^{j_{2i}} X_i^{j_{2i-1}}\bigg) | \psi \rangle 
\notag \\
\langle \psi | \bigg(\bigotimes_{i=1}^{n} X_i^{j_{2i-1}} Z_i^{j_{2i}}\bigg) ,
\end{align}

Blind quantum computation works on the commutation properties of X and Z gates such that:
\begin{equation}
    U = D \cdot U Z^b X^a,
\end{equation}

The decryption of Clifford gates does not require any interaction between the client and server during the protocol, while the non-Clifford resources require quantum interaction between the client and server to assist in the correct decryption. The decryption of these gates is given as \cite{broadbent_delegating_2015, tan_universal_2017, liu_full-blind_2020}:
\begin{align}
    H_1(X^a_1 Z^b_1 |\psi\rangle_1) &= X^b_1 Z^a_1 (H_1 |\psi\rangle_1), \label{eq:prelim_h}\\
    P_1(X^a_1 Z^b_1 |\psi\rangle_1) &= X^a_1 Z^{a\oplus b}_1 (P_1 |\psi\rangle_1), \\
   CX_{12} (X^a_1 Z^b_1 X^c_2 Z^d_2 |\psi\rangle_{12}) &= (X^a_1 Z^{b\oplus d}_1) \notag \\
                                                        & \quad (X^{a \oplus c}_2 Z^d_2) (CX_{12} |\psi\rangle_{12}),\\
   CZ_{12} (X^a_1 Z^b_1  X^c_2 Z^d_2 |\psi\rangle_{12}) &= (X^a_1 Z^{b\oplus c}_1) \notag \\
                                                        & \quad (X^c_2 Z^{a \oplus d}_2)(CZ_{12} |\psi\rangle_{12}), \label{eq:prelim_cz}
\end{align}
\begin{align}
    T_1(X_1^a Z_1^b |\psi\rangle_1 S_2^y Z_2^d | +\rangle_2) &= S_2^{a \oplus y} X_2^{a \oplus m} \notag \\
                                                            & \quad Z_2^{a(m \oplus y \oplus 1) \oplus b \oplus d \oplus y} T_1|\psi\rangle_2,
    \label{eq:t_gate_dec}
\end{align}
\begin{align}
    CCX_{123} (X^a_1 Z^b_1 X^c_2 Z^d_2  X^e_3 Z^f_3 |\psi\rangle_{123}) &= (CX^c_{13}  X^a_1 Z^b_1) \notag \\
                                                                        & \quad (CX^a_{23}  X^c_2 Z^d_2) \notag \\ 
                                                                        & \quad (CZ^f_{12} X^e_3 Z^f_3) \notag \\
                                                                        & \quad (CCX_{123}|\psi\rangle_{123}),
\end{align}

The universality of such a protocol comes from using a universal set that hides the computation from the server.
Liu et al. used a combination of $(H,S,CX,CZ,CCZ)$ \cite{liu_full-blind_2020}. Zhang et al. used the gate as a multiple of $\pi/4$ \cite{zhang_single-server_2018}.
However, these protocols based on non-parametric gates have an inherent drawback for hybrid quantum-classical algorithms, which are inherently based on parametric gates. 

\subsection{Decryption of arbitrary $R_z$ gate}

Ref. \cite{joshi2025quantum} proposed a technique of recursively decryption $R_z(\theta)$ using at most $\mathcal{O}(log_2^2(\pi/\epsilon))$ rotation gates for $\epsilon$ precision. 
Any arbitrary $\theta$ can be represented with approximate precision $\epsilon$ using $M+1$ bits where $M = \lceil log_2(\pi/\epsilon)\rceil$ using :
% The protocol decryption an arbitrary $\theta$  with approxiate precision $\epsilon$:
\begin{equation}
   \theta \approx  p_o\pi + \sum_{m=1}^M \frac{p_m\pi}{2^m},
   \label{eq:approx_theta}
\end{equation}
which can then be implemented with a series of $R_z$ gates as:
\begin{align}
    R_z(\theta) &\approx R_z(p_o\pi) \prod_{m=1}^M R_z(p_m \pi/2^m).
     \label{eq:rz_decomposed_na}
\end{align}
Here, $R_z(p_o\pi)$ can be implemented by client using $Z$ gate:
\begin{equation}
    R_z(p_o\pi) = \begin{cases}
        Z & \text{if }p_o \equiv 0 \text{ (mod 2)}, \\
        I & \text{if }p_o \equiv 1 \text{ (mod 2)}. \\
    \end{cases}
    \label{eq:rz_mpi}
\end{equation}
For $R_z(\pi/2^m)$ gates, a recursive decryption technique is used as:
\begin{equation}
    R_z(\theta)Z^bX^a = R_z(2\theta)^aX^aZ^bR_z(\theta).
    \label{eq:rz_theta_dec}
\end{equation}
This recursion stop at the base condition of $R_z(\pi/2)$ which can be decrypted using:
\begin{align}
    R_z(\pm\pi/2)X^aZ^b &= e^{\mp ia\pi/2}X^a Z^{a \oplus b}  R_z(\pm \pi/2).
\label{eq:rz_pi_2_2}
\end{align}

This procedure for recursive decryption of $R_z(\pi/2^m)$ is sketched in Algorithm \ref{algo1:dec_rz_integral}. The symbolic variable \textit{run\_of\_one} represents the availability of a subsequent $1$ in the encryption key $a$. The \textit{run\_of\_one} $=1$ implies the swap between working and ancilla qubit was not required, and \textit{run\_of\_one} implies swap was performed and needed to be corrected at the end of computation. 
This protocol utilizes at most $\mathcal{O}(log_2^2(\pi/\epsilon))$ communication rounds and $\mathcal{O}(log(\pi/\epsilon))$ steps asymptotically. Hence, it is better than any blind technique based on non-parametric gates.
However, this protocol does not hide the value of $\theta$ while delegation, hence it does not provide full-blindness to the algorithm.

\begin{algorithm}
\caption{Decryption of $R_z(\pm\pi/2^m)$ where $m \in \mathbb{Z^{+}}$}
\label{algo1:dec_rz_integral}

  \KwIn{ $\ket{\psi}$, $\theta=\pi/2^m$ where $m \in \mathbb{Z}^+$.}
  \KwResult{Decrypted state $R_z(\theta)\ket{\psi}$}
  $\ket{\phi}$ is the ancilla qubit\;
  $a_i, b_i \in_r \{0,1\}$ $\forall i \in [0, m)$\;

  $\ket{\psi} \gets Z^{b_o} X^{a_o}\ket{\psi} $\;

    \textbf{Client}$ \xrightarrow[\theta]{\ket{\psi}}$ \textbf{Server}\;

    \textbf{Server:} $\ket{\phi} \gets R_z(\theta)\ket{\psi}$ \;

    \textbf{Client}$ \xleftarrow{\ket{\psi}}$ \textbf{Server}\;

    $\ket{\psi} \gets X^{a_o}Z^{b_o}\ket{\psi}$\;

    \If{$a_o=1$}{
    $\theta \gets 2\theta$ \;
    \textit{run\_of\_one} $\gets 1$\;
    }

    \For{ $k \gets 1$ to $m$}{
     \If{ \textit{run\_of\_one} $=1$ \textbf{and} $a_{k-1} = 0$}{
     Swap($\ket{\psi} \otimes \ket{\phi}$)\;
     \textit{run\_of\_one} $\gets 2$\;
     }
     $\ket{\psi} \gets Z^{b_k} X^{a_k}\ket{\psi}$\;

    \textbf{Client}$ \xrightarrow[\theta]{\ket{\psi}}$ \textbf{Server}\;

    \textbf{Server:} $\ket{\phi} \gets R_z(\theta)\ket{\psi}$ \;

    \textbf{Client}$ \xleftarrow{\ket{\psi}}$ \textbf{Server}\;

    $\ket{\psi} \gets X^{a_k} Z^{b_k} \ket{\psi}$\;

    \If{\textit{run\_of\_one} $=2$}{
     Swap($\ket{\psi} \otimes \ket{\phi}$)\;
    }
    }
\end{algorithm}

\section{Proposed Protocol}\label{sec:proposed}
In this section, we first propose a technique of recursive decryption of an arbitrary $R_z(\theta)$ gate without revealing the value of $\theta$ and then propose a scheme of universal blind quantum computation using this technique.

In overview, this protocol extends the half-blind quantum computation protocol presented in Ref. \cite{joshi2025quantum} by proposing a technique of blindness for the communicating $\theta$. The universal computation is performed over the four-qubit resource state  $J(\epsilon)=H_1CZ_{2,3}R_z(\upsilon)_4$ where $\upsilon= p_o\pi + (\pi - \pi/2^M)$ and $M=\lceil \log_2(\pi/\epsilon)\rceil$. As this angle $\upsilon$ only depends on the precision of computation $\epsilon$, it does not reveal anything about the algorithm.

\subsection{Blind Decryption of $R_z(\theta)$}\label{subsec:blind_rz}

The protocol of recursive decryption proposed in Ref. \cite{joshi2025quantum} delegates the $\theta$ using its approximation $R_z(p_o\pi + \sum_{m=1}^M p_m\pi/2^m )$. 
This process inadvertently reveals the value of $p_m$, which in turn reveals $\theta$. 

For blind implementation of $R_z(\theta)$, we need to ensure that the server implements the gates without the knowledge of $p_m$ values. This is done by adding a strategic impurity $\eta$ to the $\theta$ before delegating it to the server, where:

\begin{align}
    \eta &= \sum_{m=1}^M (1-p_m) \frac{\pi}{2^m}.
\end{align}
This impurity is chosen such that the sumation of $\theta$ and $\eta$ becomes a constant, as (see proof, Appendix \ref{app:upsilon_indep}):
\begin{align}
    \upsilon &= \theta + \eta, \notag \\
            &= p_o\pi + \bigg(\pi - \frac{\pi}{2^M}\bigg).
\end{align}

As this value is independent of $p_m$, delegating $\upsilon$ does not reveal anything about the $\theta$. 
Note, $\upsilon$ is dependent on $M=\lceil \log_2(\pi/\epsilon)\rceil$, which is defined by the precision of computation $\epsilon$ and is independent of the type of algorithm running on the server. 

The protocol then uses $H$, $X$, and $Swap$ gates to selectively decrypt $R_z(\theta)$ at the client side.
We start by representing $\upsilon$ using a geometric sequence of $\pi/2^m$ elements as:
\begin{align}
    R_z(\upsilon) &= R_z\bigg(p_o\pi  + \sum_{m=1}^M \pi/2^m\bigg), \notag \\
                  &= R_z(p_o\pi) \prod_{m=1}^M R_z(\pi/2^m).
\end{align}
Here, $R_z(p_o\pi)$ can be implemented at the client side using $Z$ gate if the $p_o$ is odd; otherwise, no additional gate is required as given in Eq. (\ref{eq:rz_mpi}). The protocol then uses $Swap$ gate to decrypt the $R_z(p_m\pi/2^m)$ using the series of $R_z(\pi/2^m)$ gates. The protocol utilises one ancilla qubit $\ket{\phi}$ and performs correct decryption of $R_z(\theta)$ on working qubit $\ket{\psi}$.

For delegation of $R_z(p_m\pi/2^m)$, there are three cases as $p_m \in \{1,0,-1\}$.
A symbolic variable $s_m$ is taken such that:
\begin{equation}
    s_m = \begin{cases}
        1, & \text{ if } p_m \in \{1,-1\}, \\
        0, & \text{ if } p_m = 0.
    \end{cases}
    \label{eq:sm}
\end{equation}
If $s_m = 1$,the ancilla $\ket{\phi}$ and working qubit $\ket{\psi}$ are swapped, such that recursive decryption of $R_z(\pi/2^m)$ have an effect on the working qubit only when $p_m \in \{1,-1\}$.  However, this leaves the decryption when $p_m=-1$ incorrect.
For this, another symbolic variable $q_m$ is taken such that:
\begin{equation}
    q_m = \begin{cases}
        1, & \text{ if } p_m = -1, \\
        0, & \text{ if } p_m = 1.
    \end{cases}
    \label{eq:qm}
\end{equation}
If the value of $q_m=1$, then this requires an additional condition on the swap that is applied to ensure the correct decryption on the working qubit based on the following (see proof, Appendix \ref{eq:approx_theta}):
\begin{equation}
    R_z((-1)^q\theta) Z^b X^a = R_z(2\theta)^{a \oplus q} Z^b X^a R_z((-1)^{q}\theta).
    \label{eq:minus_rz_theorem}
\end{equation}
Algorithm \ref{algo3:ubqc} sketches the complete protocol of recursive decryption of $R_z(\theta)$ without revealing the value of $\theta$. Here, the notation \textbf{Client}$ \xrightarrow[c]{q}$ \textbf{Server} is used to denote a communication channel between client and server, where $q$ denotes quantum information and $c$ denotes classical information.
The total communication rounds between the client will be at most $M^2=\mathcal{O}(\log_2^2(\pi/\epsilon))$. The inner loop can be optimised with an early breaking condition, which will result in asymptotic complexity of $\mathcal{O}(\log_2(\pi/\epsilon))$.

\begin{algorithm}
 \caption{Blind Decryption of $R_z(\theta)$}
 \label{algo3:ubqc}

\KwIn{ $\ket{\psi}$, $\theta$, $\epsilon$}
\KwResult{$R_z(\theta)|\psi\rangle$}

$M \gets \lfloor \log_2(\pi/\epsilon) \rfloor$\;
$p \gets \lfloor \theta/\pi \rfloor $ \;
$ \ket{\psi_c} \gets \ket{\phi} \otimes \ket{\psi}$ \;

\If{ $p \equiv 1 \text{ (mod 2)} $}{
$\ket{\psi} \gets \ket{\phi} \otimes Z\ket{\psi}$ \;
}

\For{$m \gets 1$ to $M$}{
$ p_m \gets \lfloor 2^m ((\theta-p)/\pi)\rfloor - 2 \lfloor 2^{m-1} ((\theta-p)/\pi)\rfloor$ \;
$
    s_m \gets \begin{cases} 1, & \text{if } p_m \in \{1,-1\} \\ 
                      0, & \text{if } p_m = 0 \end{cases}
$\;

$
 q_m \gets \begin{cases}
     1, & \text{if } p_m = -1 \\
     0, & \text{if } p_m = 1 \\ 
 \end{cases}
$\;

$\ket{\psi_c} \gets Swap^{s_m} (\ket{\phi} \otimes \ket{\psi})$ \;

\For{$k \gets m$ to $1$}{

$a_k, b_k \in_r \{0,1\}$\;

$\ket{\phi} \gets Z^{b_k} X^{a_k} \ket{\phi}$\;

\textbf{Client}$ \xrightarrow[k]{\ket{\phi}}$ \textbf{Server}\;

\textbf{Server:} $\ket{\phi} \gets R_z(\pi/2^k)\ket{\phi}$ \;

\textbf{Client}$ \xleftarrow{\ket{\phi}}$ \textbf{Server}\;

\If{$k=1$}{
$\ket{\phi} \gets X^{a_1} Z^{b_1 \oplus a_1} \ket{\phi}$\;
$\ket{\psi_c} \gets Swap^{s_m  \cdot\prod_{i=2}^{m} (a_i \oplus q_m) } (\ket{\phi} \otimes \ket{\psi})$\;
}

\Else{
$\ket{\phi} \gets X^{a_k} Z^{b_k} \ket{\phi}$\;
$\ket{\psi_c} \gets Swap^{s_m \cdot (\bar{a}_k \oplus \bar{q}_m)\cdot \prod_{i=k-1}^{m} (a_i \oplus q_m)  } (\ket{\phi} \otimes \ket{\psi})$\;
}
}
}
\label{algo2:blind_rz}
\end{algorithm}

\subsection{Universal Blind Quantum Computation}

In this subsection, we show how the blind decryption of $R_z(\theta)$ can be used to perform universal blind quantum computation. 

The proposed protocol requires a client capable of performing gates from the set $\mathcal{C} \in \{X, Z, Swap, Measure\}$ and a server that needs the capability to perform gates from the set $\mathcal{S} \in \{H, CZ, R_z\}$ to perform universal computation. The protocol requires both a quantum and a classical channel between clients. 

\begin{figure*}
    \centering
    \includegraphics[width=0.9\linewidth]{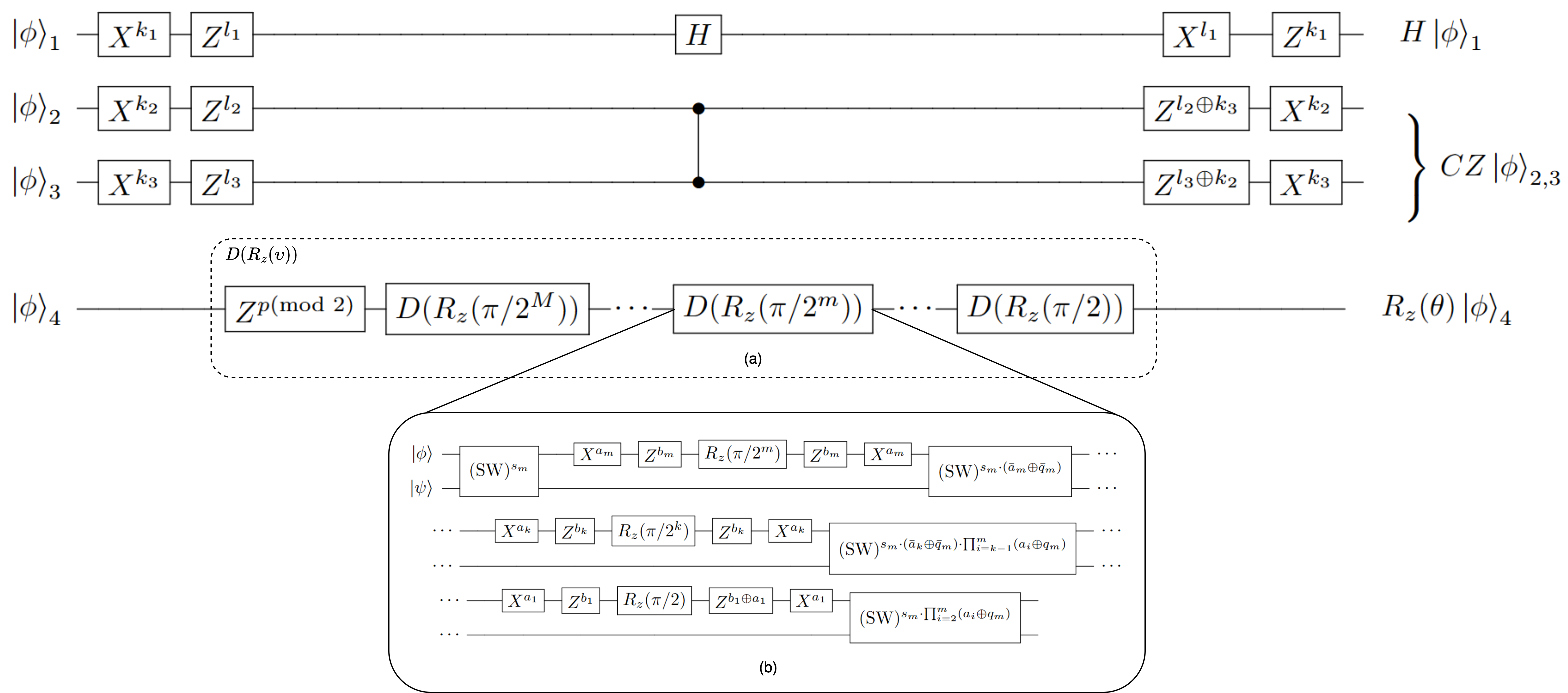}
    \caption{Resource set $J(\epsilon) =H_1CZ_{2,3}R_z(\upsilon)_4$ for Universal Blind Quantum Computation, where (a) represents recursive decryption of $R_z(\upsilon)$ using representation $D(R_z(\pi/2^m))$ $\forall m\in \{1,\cdots,M\}$. (b) represents the exact procedure of recursive decryption using encryption keys $a_i,b_i \in_r \{0,1\}$ $\forall i \in \{1, \cdots, m\}$ and boolean variables $s_m$ and $q_m$ such that $p_m = (-1)^{q_m}s_m$ for blind decryption of $R_z(\theta) = R_z(p_o\pi)\prod_{m=1}^M R_z(p_m\pi/2^m)$. }
    \label{fig:ubqc_resource_set}
\end{figure*}

The universal computation is performed over a four-qubit universal resource set $J(\epsilon) = H_1CZ_{2,3}R_z(\upsilon)_4$. As this resource set does not depend on $\theta$, the server will be completely blind to the type of computation being performed. 
Here, gates $H$ and $CZ$ do not require any interaction. $R_z(\upsilon)$ gate requires $\mathcal{O}(log_2^2(\pi/\epsilon))$ communication rounds between client and server. Fig. \ref{fig:ubqc_resource_set} shows the delegation of resource set $J(\epsilon) =H_1 CZ_{2,3} R_z(\upsilon)_4 $ where $\upsilon = \pm \pi(1-1/2^M)$, and $M=\lceil log_2(\pi/\epsilon) \rceil$.
To simplify the notation, we have denoted $Swap$ gates as $SW$ in the figure.
The Fig. \ref{fig:ubqc_resource_set}(a) showcase the procedure of recursive decryption of $R_z(\upsilon)$ denoted by $D(R_z(\upsilon))$. Fig. \ref{fig:ubqc_resource_set}(b) denotes the decryption of each individual $R_z(\pi/2^m)$ using conditional $Swap$ gates.

The procedure of universal blind quantum computation using $J(\epsilon)$ is performed as:
Client starts by creating a computation set $\mathcal{J} = \{U_{g,q} \mid g \in \{1,...,n_g\}\}$ where  $q$ is ordered set of qubit on which $U_g$ is implemented, from a given algorithm $\mathcal{A}$ with $n$ qubits and $n_g$ number of gates.

If the gate in the computation set $\mathcal{J}$ belongs to the gate in the client set $\mathcal{C}$, then the client performs the gate itself; otherwise, it is delegated to the server. 

For delegation of the gate to the server, the client prepares a four-qubit ancilla set $\ket{\phi}$ which will be transmitted between the client and server.
If the gate to be delegated is the $H$ gate, then the client swaps the first qubit of the ancilla set $\ket{\phi}_1$ with the working qubit $\ket{\psi}_q$. If the gate to be delegated is $CZ$, the client swaps the second and third qubits of the ancilla set $\ket{\phi}_{2,3}$ with the working qubits $\ket{\psi}_q$. If the gate to be delegated is $R_z$, the client swaps the fourth qubit of the ancilla set $\ket{\psi}_4$ with the working qubit $\ket{\psi}_q$.

The client encrypts the outgoing ancilla state using $k_i,l_i \in_r \{0,1\}$ $\forall i \in \{0,1,2\}$ as:
\begin{equation}
    \ket{\phi}_{1,2,3} \gets \bigg( \bigotimes_{i=1}^3 Z_i^{l_i} X_i^{k_i}\bigg) \ket{\phi}_{1,2,3}.
\end{equation}
The $\ket{\phi}_4$ is encrypted implicitly during the recursive decryption of $R_z(\theta)$ using at most $2log_2(\pi/\epsilon)$ keys. 

The client then sends the state $\ket{\phi}$ to the server, which then implements the operation $J(\epsilon)$ where $H$ is implemented on the first qubit, $CZ$ is implemented on the second and third qubit. Server performs recursive decryption of $R_z(\theta)$ interactively using Algorithm \ref{algo2:blind_rz} as described in Sec. \ref{subsec:blind_rz}. 

The server then send this state $\ket{\phi}$ back to client, who then decrypts the remaining qubits $\ket{\phi}_{1,2,3}$ as:
\begin{equation}
    \ket{\phi}_{1,2,3} = Z_3^{k_3 \oplus l_2} X_3^{l_3} Z_2^{k_2 \oplus l_3} X_2^{l_2}X_1^{k_1}Z_1^{l_1}\ket{\phi}_{1,2,3}.
\end{equation}
The client then swaps back the working and ancilla qubit according to the given gates.
The procedure is sketched in Algorithm \ref{algo3:hdqc_algo}.

\begin{algorithm}
\caption{Proposed Full-Blind Quantum Computation Protocol}
\label{algo3:hdqc_algo}

\KwIn{Algorithm $\mathcal{A}$, total qubits $n$, total number of gate $n_g$, precision $\epsilon$.}

$\upsilon \gets \pi(1-1/2^{\lceil log_2(\pi/\epsilon) \rceil}) $\;
Client set $\mathcal{C} \in \{X,Z,Swap,Measure\}$ \;
Server set $\mathcal{S} \in (H_1, CZ_{2,3}, R_z(\upsilon)_4)$ \;
Create $\mathcal{J} \gets \{U_{g,q} \mid g \in \{0, \dots, n_g\}, q \text{ is ordered set of qubits}\}$ \;
$J(\epsilon) = R_z(\upsilon)_4 CZ_{2,3} H_1$\;
$\ket{\phi}$ is a four-qubit ancilla state \;

\For{$j \gets 1$ \KwTo $n_g$}{
 \If{ $U_{j,q} \in \mathcal{C}$}{
    $\ket{\psi}_q \gets U_{j,q}\ket{\psi}_q$\;
  }
  \Else{
    \If{$U_{j,q}= H$}{
    $Swap(\ket{\phi}_1 \otimes \ket{\psi}_q)$
    }
    \ElseIf{$U_{j,q}= CZ$}{
    $Swap(\ket{\phi}_{2,3} \otimes \ket{\psi}_q)$
    }
     \ElseIf{$U_{j,q}= R_z$}{
    $Swap(\ket{\phi}_4 \otimes \ket{\psi}_q)$
    }
    
    $k_i,l_i \in_r \{0,1\}$ $\forall i \in \{1,2,3\}$\;
  $\ket{\phi}_{1,2,3} \gets \Big(\bigotimes_{i=1}^{3} Z_i^{l_i} X_i^{k_i} \Big) \ket{\phi}_{1,2,3} $\;

  \textbf{Client}$\xrightarrow{\ket{\phi}}$ \textbf{Server}\;
  \textbf{Server:} $\ket{\phi} \gets J(\epsilon)\ket{\phi}$ interactively using Algorithm \ref{algo2:blind_rz} \;
  \textbf{Client}$ \xleftarrow[]{\ket{\phi}}$ \textbf{Server}\;
  
  $\ket{\phi}_{1,2,3} \gets Z_3^{k_3 \oplus l_2} X_3^{l_3} Z_2^{k_2\oplus l_3} X_2^{l_2}X_1^{k_1}Z_1^{l_1}\ket{\phi}_{1,2,3}$ \;  

    \If{$U_{j,q}= H$}{
    $Swap(\ket{\phi}_1 \otimes \ket{\psi}_q)$
    }
    \ElseIf{$U_{j,q}= CZ$}{
    $Swap(\ket{\phi}_{2,3} \otimes \ket{\psi}_q)$
    }
     \ElseIf{$U_{j,q}= R_z$}{
    $Swap(\ket{\phi}_4 \otimes \ket{\psi}_q)$
    }

  }
}
\end{algorithm}

\textbf{Proof of Universality:}
A gate set like $\{X,Z,H,S,T,CX\}$ is considered to be universal for quantum computation \cite{nielsen2001quantum}. It follows directly that a client restricted to implementing only $X$ and $Z$ gates can perform universal computation with the help of a server capable of performing gates from the set $\{H,CZ,R_z\}$.

With the help of Euler's decomposition, such a gate set can represent any single qubit operation using $R_z$ and $R_x$ as:
\begin{equation}
    U = e^{i\phi} R_z(\alpha)R_x(\beta)R_z(\gamma),
\end{equation}
and using the identity $R_x(\theta) = H R_z(\theta)H $ any single-qubit gate can be performed using $H, R_z(\theta)$.

Moreover, the identity $CX_{1,2} = H_2CZ_{1,2}H_2$ implies that controlled-NOT can be generated from operations available at the server's end. Together with the non-Clifford resource $T(=R_z(\pi/4))$, this suffices to construct an arbitrary multi-qubit operation.

\textbf{Proof of Correctness:}
For proof of correctness, we need to show that the client's view of the protocol is what the client needed to implement. 
For the universal resource set $J(\epsilon)= H_1CZ_{2,3}R_z(\upsilon)_4$, it is trivial to show the correctness of the $H$ and $CZ$ gate using the equivalence rule of gate decryption given in Eq. (\ref{eq:prelim_h}) and Eq. (\ref{eq:prelim_cz}).
The correctness of $R_z(\theta)$ as implementated using Algorithm \ref{algo2:blind_rz} can be shown as:
The circuit in Fig. \ref{fig:ubqc_resource_set}(b) is equivalent to $R_z((-1)^{q_m}s_m \pi/ 2^m)$ (see proof, Appendix \ref{app:swap_equiv}). 
Using recursive decryption for all the values of $m \in \{1, \cdots, M\}$, we can represent the delegation of $\theta'$ as:
\begin{align}
    R_z(\theta') &= \prod_{m=1}^M R_z((-1)^{q_m} \pi/2^m)^{s_m}, \notag \\
                 &= \prod_{m=1}^M R_z((-1)^{q_m} s_m \pi/2^m). \label{eq:rz_dash}  
\end{align}
Note, we have used the following identity associated with the boolean variable $s_m$:
\begin{equation}
    R_z(\theta')^{s_m} = R_z(s_m\theta').
\end{equation}
Based on the values of boolean variables $s_m$ and $q_m$, we can represent $p_m = (-1)^{q_m}s_m \in \{1,0,-1\}$. This make Eq. (\ref{eq:rz_dash}) equivalent to:
\begin{align}
    R_z(\theta') = \prod_{m=1}^M R_z(p_m \pi/2^m).
\end{align}
As the value $p_o\pi/2^m$ can be implemented by the client using the $Z$ gate only. Hence, the value can be $R_z(\theta)$ can be correctly delegated as:
\begin{align}
    R_z(\theta) &= R_z(p_o\pi + \theta'), \notag \\
                &= R_z(p_o\pi)\prod_{m=1}^M R_z(p_m \pi/2^m).
\end{align}
Hence, proving the correctness of the scheme.

\textbf{Proof of Blindness:}
For proof of blindness, we need to show that $J(\epsilon)$ is independent of the algorithm $\mathcal{A}$. We first start by proving that the recursive decryption of $R_z(\theta)$ using $R_z(\upsilon)$ is blind. 

During the execution of Algorithm \ref{algo2:blind_rz}, quantum information $\ket{\phi}$ and classical information $k$ are revealed to the server. 
Ref. \cite{joshi2025quantum} showed that quantum information $\ket{\phi}$ transmitted during recursive decryption is blind using an entanglement-equivalent circuit. However, the value of $k$ revealed the rotation angle $\theta$.
Here, server perform rotation of $\upsilon$ instead of $\theta$, so using the value of $k$ server gets to know that 
\begin{align}
    \upsilon 
             % &= p_o \pi + \bigg( \sum_{m=1}^M \frac{\pi}{2^m} \bigg),\\ 
            &= p_o\pi + \bigg(\pi - \frac{\pi}{2^M}\bigg).
\end{align}
As the value of $\upsilon$ is only dependent on $M=\lceil\log_2(\pi/\epsilon)\rceil$, which is a function of precision $\epsilon$ (see proof, Appendix \ref{app:upsilon_indep}). This does not reveal anything about the algorithm, i.e., $P(\mathcal{A}|J(\epsilon),\theta)= P(\mathcal{A}|\theta)$.
Hence, we can use this to prove that the protocol does not change the server's belief about the nature of the algorithm:
\begin{align}
    P(J(\epsilon)|\mathcal{A},\theta) &= \frac{P(\mathcal{A}|J(\epsilon),\theta) P(J(\epsilon))}{P(\mathcal{A}|\theta)}, \notag\\
                    &= P(J(\epsilon)).
\end{align}
This shows that the server's belief about the nature of the algorithm is not changed, i.e., no information about $A$ is revealed to the server. Hence, the protocol is blind.

\section{Comparative Analysis}\label{sec:compartive_analysis}
Let $n_p:n_{np}$ be the ratio between parametric and non-parametric gates in a given algorithm $\mathcal{A}$.
Then the quantum communication cost $C_p$ incurred by algorithms capable of decryption only parametric gates is given by:
\begin{equation}
    C_p = n_{p} \ln^{3.97}(1/\epsilon) + n_{np} \cdot 1.
\end{equation}
While the proposed protocol requires a constant $\mathcal{O}(log_2^2(\pi/\epsilon))$ for all gates, hence communcation cost $C_{np}$ will be given as:
\begin{equation}
    C_{np} = (n_{np} + n_{p}) \log_2^2(\pi/\epsilon).
\end{equation}

\begin{figure}
    \centering
    \includegraphics[width=0.9\linewidth]{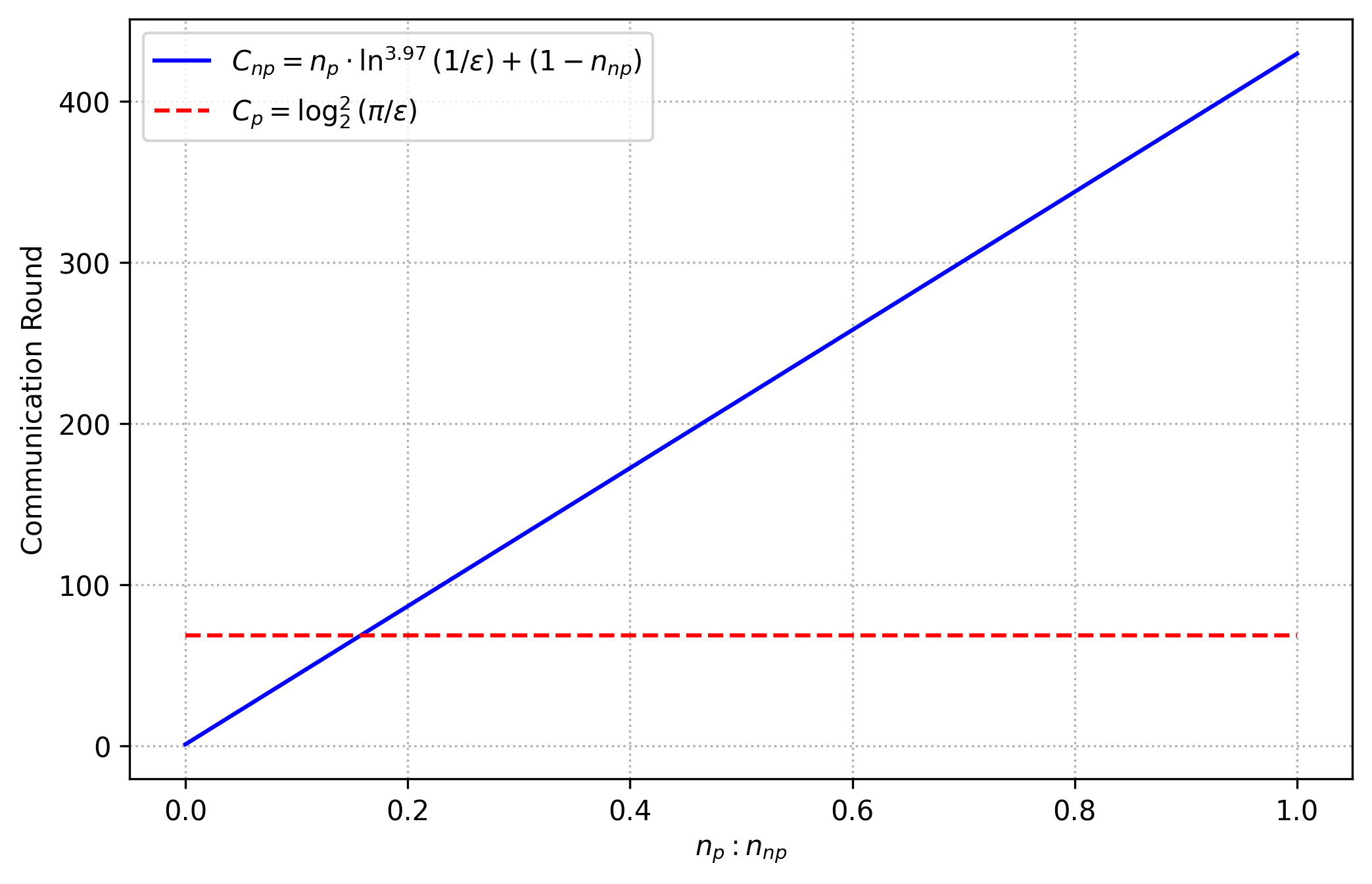}
    \caption{Comparison of communication cost between protocols using only parametric gates $C_p$ using Solovay-Kitaev decomposition vs the proposed protocol $C_{np}$ that can inherently decrypt non-parametric gates without prior decomposition. 
    % Here $n_p:n_{np}$ is the ratio of parameter to non-parametric gates present in a given algorithm. 
    Here $\epsilon=10^{-2}$. }
    \label{fig:intersection}
\end{figure}

\begin{figure}
    \centering
    \includegraphics[width=0.9\linewidth]{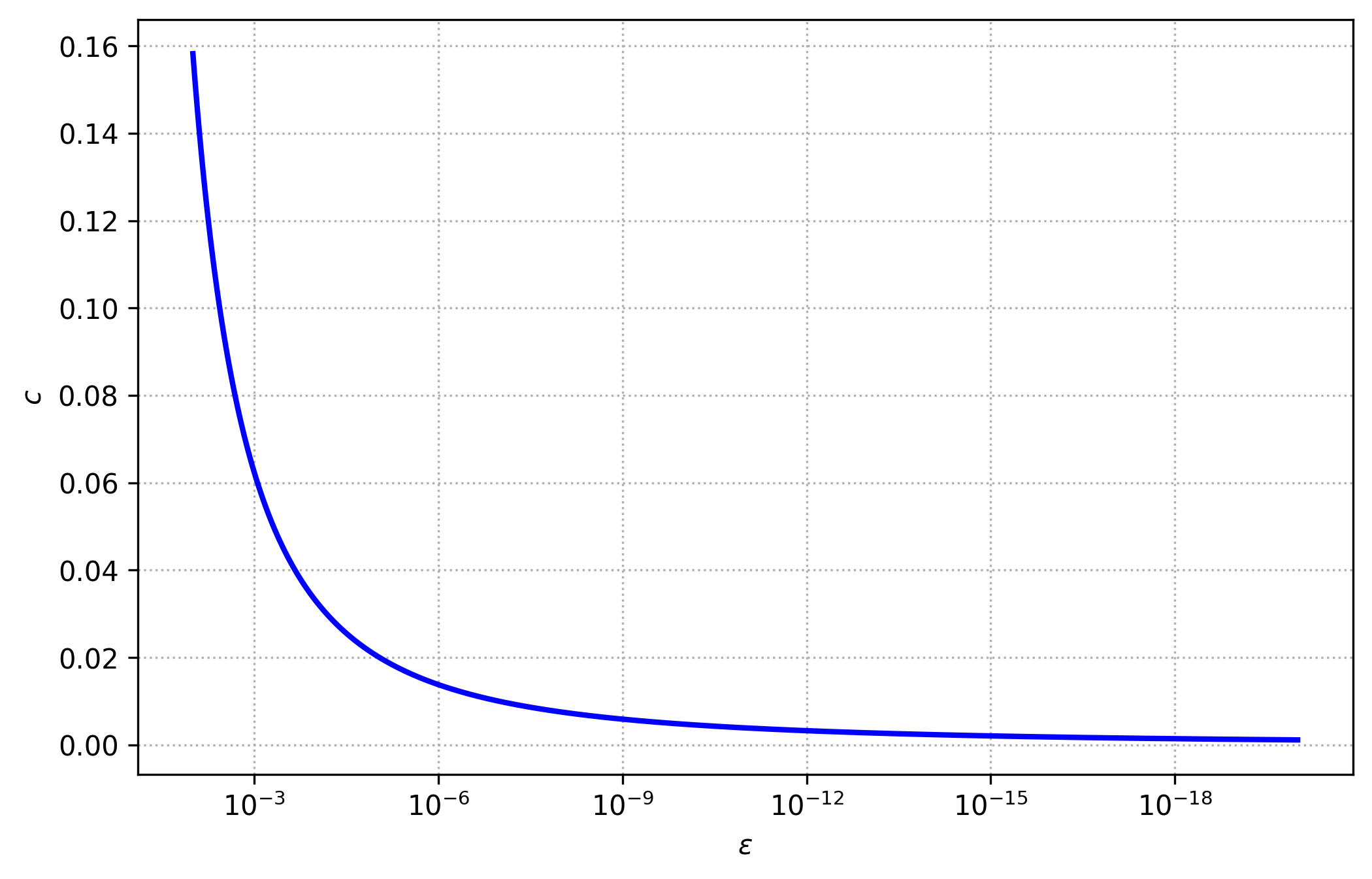}
    \caption{The plot of critical ratio $c$ needed for the proposed protocol to require fewer communication rounds than the protocols based on parametric resources only with respect to the precision of computation $\epsilon$.}
    \label{fig:critical_ratio}
\end{figure}

Fig. \ref{fig:intersection} gives the comparison between $C_p$ and $C_{np}$ as the precision of the computation is increased for $\epsilon=10^{-2}$. 
This shows that the proposed protocol becomes profitable only if the given algorithm has a ratio of $n_p:n_{np}$ higher than a certain value, let's say the critical ratio $c$. This point can be defined using:
\begin{align}
    c &= \frac{log_2^2(\pi/\epsilon) - 1}{ln^{3.97}(1/\epsilon)-1},
\end{align}
such that $\epsilon \neq 1/e $.
Fig. \ref{fig:critical_ratio} shows how this critical ratio decreases with the increase in precision $\epsilon$ demanded. 
For a particular value of $\epsilon=10^{-10}$, we get the value of $x = 0.005$, which implies that out of $1000$ gates, if only $4$ gates are parametric, the algorithm will be profitable. 

\section{Conclusion}\label{sec:conclusion}
In this paper, we have extended the protocol of half-blindness proposed in Ref. \cite{joshi2025quantum} by delegating the $R_z(\upsilon)$ gate instead of $R_z(\theta)$. Here, $\upsilon=p_o\pi + \pi - \pi/2^{\log_2(\pi/\epsilon)}$, which depends only on the precision of computation and not the algorithm running on the server. The proposed protocol then extracts the correct $R_z(\theta)$ at the client's side using $\{X,Z,Swap\}$ gates. This technique has been then utilized to propose a universal blind quantum computing protocol using resource set $J(\epsilon)= H_1CZ_{2,3}R_z(\upsilon)_4$. The protocol requires at most $\mathcal{O}((n_p+n_{np})\log_2^2(\pi/\epsilon))$ steps with asymptotic complexity of $\mathcal{O}((n_p + n_{np})\log_2(\pi/\epsilon))$. Here $n_p:n_{np}$ is the ratio of parametric to non-parametric gates in a given algorithm. As this is the first protocol capable of decrypting parametric gates, it does not require the prior decomposition of gates as required by non-parametric resources-based protocols, which incurs a computation complexity of $\mathcal{O}(n_{p}\ln^{3.97}(1/\epsilon) + n_{np})$, using Solovay-Kitaev decomposition.

\begin{acknowledgments}
This research is supported by a seed grant under the IoE, BHU [grant no. R/Dev/D/IoE/SEED GRANT/2020-21/Scheme No. 6031].
\end{acknowledgments}

\appendix

\section{$\upsilon$ is independent of $\theta$}\label{app:upsilon_indep}

For a given $\epsilon$-approximation of $\theta=p_o\pi + \sum_{m=1}^M p_m \pi/2^m$, we take a strategic impurity $\eta$ such that:
\begin{align}
    \eta = \sum_{m=1}^M (1-p_m)\frac{\pi}{2^m}
\end{align}
where $M=\lceil \log_2(\pi/\epsilon)\rceil$.
Then the summation $\upsilon$ of $\theta$ and $\eta$ will be:
\begin{align}
    \upsilon &= \theta + \eta, \notag\\
             &= p_o\pi + \sum_{m=1}^M p_m \frac{\pi}{2^m} + \sum_{m=1}^M (1-p_m)\frac{\pi}{2^m}, \notag\\
             % &= p_o\pi + \sum_{m=1}^M p_m \frac{\pi}{2^m} + \sum_{m=1}^M \frac{\pi}{2^m} - \sum_{m=1}^M p_m \frac{\pi}{2^m} \\
             &= p_o\pi + \sum_{m=1}^M \frac{\pi}{2^m}.
\end{align}
Using the sumation of geometric series $\pi/2^m$, where first term and common ratio is $\pi/2$, we get:
\begin{align}
    \upsilon &= p_o\pi + \pi - \frac{\pi}{2^M}.
\end{align}
As $M$ is only dependent on the precision of computation $\epsilon$, the $\upsilon$ becomes independent of $\theta$ needed for the algorithm running on the server.

\section{Decryption of arbitrary $z$ gate}

\begin{theorem}\label{thrm1:rz_gate}
The decryption of arbitrary $R_z(\eta)$ where $\eta = (-1)^q \theta$ is dependent on $R_z(2\theta)$, i.e.,
$R_z(\eta)Z^bX^a = R^{a \oplus q}_z(2\theta)X^aZ^bR_z(\eta)$.
\end{theorem}

\begin{proof}
To perform decryption of $R_z$ gate under Pauli's $X$ and $Z$ encryption, we need to determine the value of unitary $D$ such that:
% For decryption of $R_z$ gate encrypted by $X$ and $Z$ gate, we need to find a unitary $D$ such that:
\begin{equation}
    R_z(\theta) = D \cdot R_z(\eta) Z^b X^a,
\end{equation}
where $a,b \in_r \{0,1\}$ and $\eta = (-1)^q \theta$.
Solving for $D$, we obtain:
% Here, solving for $D$ we get:
\begin{align}
    D &= R_z(\theta) X^a Z^b R_z(-\eta).
\end{align}

Note that $R_z^\dagger(\theta) = R_z(-\theta)$.
Also,
\begin{equation}
    R_z(\theta) = \begin{pmatrix}
        e^{-i\theta /2} & 0 \\
        0 & e^{i\theta/2}
    \end{pmatrix}.
\end{equation}

Since $X$ and $Z$ are standard Pauli operations, the binary power of these operators can be expressed as:
% Given $X$ and $Z$ are standard Pauli rotation gates, we can represent $X^a$ and $Z^b$ as:
\begin{align}
    X^a = \begin{pmatrix}
        1-1_a & 1_a \\
        1_a & 1-1_a \\
    \end{pmatrix}, &
    Z^b = \begin{pmatrix}
        1 & 0 \\
        0 & (-1)^{1_b}
    \end{pmatrix},
\end{align}

where 
\begin{equation}
1_x = \begin{cases} 1, & \text{if } x =1, \\ 
0, & \text{if } x=0.
\end{cases}
\end{equation}

Using indicator variable formulation of $X$ and $Z$ gates, we can perform algebraic manipulation to matrix $D$, giving us the following representation: 
\begin{equation}
    D = \begin{pmatrix}
        (1-1_a)e^{i(\eta-\theta)/2} & (-1)^{1_b} 1_a e^{i(-\eta -\theta)/2} \\
        1_a e^{i(\eta +\theta)/2} & (-1)^{1_b}(1-1_a)e^{i(-\eta+\theta)/2}
    \end{pmatrix}.
\end{equation}
This matrix representation of $D$ admits to further decomposition as:
\begin{align}
   D &= \begin{pmatrix}
        e^{i(\eta(-1)^{1_a}-\theta)/2} & 0 \\
        0 & e^{-i(\eta(-1)^{1_a}-\theta)/2}
    \end{pmatrix} \notag \\
    & \quad \quad 
    \begin{pmatrix}
        1-1_a & 1_a \\
        1_a & 1-1_a \\
    \end{pmatrix}
        \begin{pmatrix}
        1 & 0 \\
        0 & (-1)^{1_b}
    \end{pmatrix}, \notag\\
    &= R^{a\oplus q}_z(2\theta) X^a Z^b.
\end{align}

Here, we have used the following identities associated with indicator variables $1_a$ and $1_b$, where $a,b \in_r \{0,1\}$:
\begin{align}
    (1- (-1)^{1_a}(-1)^{1_q}) &= 2(1_a \oplus 1_q),  &  (-1)^{2\cdot 1_b}  &= 1, \notag\\
    (1-1_a)^2 &= 1-1_a,   & (1-1_a)\cdot 1_a &= 0,  \notag \\
       (1-1_a)x + 1_ay &= x^{1-1_a}y^{1_a},   &  (1-2\cdot1_a) &= (-1)^{1_a}.
\end{align}

Hence,
\begin{equation}
    R_z(\eta)Z^bX^a = R_z^{a \oplus q}(2\theta)X^aZ^bR_z(\eta).
\end{equation}
This shows that the decryption of the rotation gate $R_z(\eta)$ is dependent on the $R_z(2\theta)$ gate.
\end{proof}

\section{Equivalence of $Swap$ circuit}\label{app:swap_equiv}

\begin{figure*}
    \centering
 \[
\scalebox{0.7}{
\Qcircuit @C=0.3em @R=.7em {
% \lstick{\ket{\psi}} 
& \qw      & \gate{X^{s_m \cdot a_{m}}} & \gate{Z^{s_m \cdot b_{m}}} & \gate{R_z(\pi/2^{m})^{s_m}} & \gate{Z^{s_m \cdot b_{m}}} & \gate{X^{s_m \cdot a_{m}}} & 
\qw &
\gate{X^{s_m \cdot a_m \cdot (a_{m-1} \oplus q_m)}} & \gate{Z^{s_m \cdot a_m \cdot (b_{m-1} \oplus q_m)}} & \gate{R_z(\pi/2^{m-1})^{s_m \cdot (a_m \oplus q_m)}} & \gate{Z^{s_m \cdot a_m \cdot (b_{m-1} \oplus q_m)}} & \gate{X^{s_m \cdot a_m \cdot (a_{m-1} \oplus q_m)}} & \qw & 
\quad \cdots \quad 
}
}
\]
 \[
\scalebox{0.7}{
\Qcircuit @C=0.3em @R=.7em {
\quad \cdots \quad &   & \qw      & \gate{X^{s_m \cdot (\prod_{i=k+1}^{m} a_i ) \cdot (a_k \oplus q_m)}} & \gate{Z^{s_m \cdot (\prod_{i=k+1}^{m} a_i ) \cdot (b_k \oplus q_m)}} & \gate{R_z(\pi/2^{k})^{s_m \cdot (\prod_{i=k+1}^{m} a_i ) \cdot (a_k \oplus q_m)}} & \gate{Z^{s_m \cdot (\prod_{i=k+1}^{m} a_i ) \cdot (b_k \oplus q_m)}} & \gate{X^{s_m \cdot (\prod_{i=k+1}^{m} a_i ) \cdot (a_k \oplus q_m)}} & 
 \qw & \quad \cdots \quad
}
}
\]
\[
\scalebox{0.7}{
\Qcircuit @C=0.3em @R=.7em {
\quad \cdots \quad & &  \qw      & \gate{X^{s_m \cdot (\prod_{i=2}^{m} a_i ) \cdot (a_{1} \oplus q_m)}} & \gate{Z^{s_m \cdot (\prod_{i=2}^{m} a_i ) \cdot (b_{1} \oplus q_m)}} & \gate{R_z(\pi/2)^{s_m \cdot (\prod_{i=2}^{m} a_i ) \cdot (a_1 \oplus q_m)}} & \gate{Z^{s_m \cdot (\prod_{i=2}^{m} a_i ) \cdot (b_{1} \oplus a_1)}} & \gate{X^{s_m \cdot (\prod_{i=2}^{m} a_i ) \cdot a_{1}}} & 
\qw 
}
}
\]
=
\[
\Qcircuit @C=0.3em @R=.7em {
\lstick{\ket{\psi}} & \gate{R_z((-1)^{q_m} \pi /2^m)^{s_m}} & \qw 
}
\]
     \caption{Equivalent circuit without Swap gate for circuit given in Fig. \ref{fig:ubqc_resource_set}(b) using identies associated with $Swap$ gate. The recursive decryption is equivalent to $R_z((-1)^{q_m}\pi/2^m)^{s_m}$ using Eq. (\ref{eq:minus_rz_theorem}).}
    \label{fig:equiv_rz_gate}
\end{figure*}

Considering the following identities associated with the $Swap$ gate:

\[
\scalebox{0.6}{
\Qcircuit @C=1em @R=.7em {
 & \qw      & \multigate{1}{(\mathrm{SW})^{s_1}}      & \gate{G_1}      & \multigate{1}{(\mathrm{SW})^{s_1}}     & \gate{G_2} & \qw \\
 & \qw      & \ghost{(\mathrm{SW})^{s_1}} & \qw & \ghost{(\mathrm{SW})^{s_1}}      & \qw & \qw
}
\quad = \quad
\Qcircuit @C=1em @R=.7em {
 &  \gate{G_1^{\bar{s}_1}} & \gate{G_2^{s_1 \odot s_2}} &\multigate{1}{(\mathrm{SW})^{s_1 \oplus s_2}} & \qw\\
 &  \gate{G_1^{s_1}} & \gate{G_2^{s_1 \oplus  s_2}} & \ghost{(\mathrm{SW})^{s_1 \oplus s_2}} &\qw
}}
\]

Let's say $s_2 \gets s_1\cdot \bar{s}_2$, then using following boolean identities,
\begin{align}
    a \oplus (a\cdot \bar{b}) = a\cdot b, \\
    \bar{a} \odot \overline{(a \cdot \bar{b})} = \bar{a} + \bar{b}.
\end{align}
We can simplify the circuit above as:
\[
\scalebox{0.6}{
\Qcircuit @C=1em @R=.7em {
 & \qw      & \multigate{1}{(\mathrm{SW})^{s_1}}      & \gate{G_1}      & \multigate{1}{(\mathrm{SW})^{s_1\cdot \bar{s}_2}}     & \gate{G_2} & \qw \\
 & \qw      & \ghost{(\mathrm{SW})^{s_1}} & \qw & \ghost{(\mathrm{SW})^{s_1\cdot \bar{s}_2}}      & \qw & \qw
}
\quad = \quad
\Qcircuit @C=1em @R=.7em {
 &  \gate{G_1^{\bar{s}_1}} & \gate{G_2^{\bar{s}_1 + \bar{s}_2}} &\multigate{1}{(\mathrm{SW})^{s_1 \oplus s_2}} & \qw\\
 &  \gate{G_1^{s_1}} & \gate{G_2^{s_1 \cdot  s_2}} & \ghost{(\mathrm{SW})^{s_1 \oplus s_2}} &\qw
}}
\]
For a setup of $k$ $Swap$ gate given below:
\[
\scalebox{0.6}{
\Qcircuit @C=1em @R=.7em {
 & \qw      & \multigate{1}{(\mathrm{SW})^{s_1}}      & \gate{G_1}      & \multigate{1}{(\mathrm{SW})^{s_1\cdot \bar{s}_2}}  & \gate{G_2} & \qw & \quad \cdots \quad && \multigate{1}{(\mathrm{SW})^{(\prod_{i=1}^{k-1} s_1)\cdot \bar{s}_k}} & \gate{G_k} & \qw \\
 & \qw      & \ghost{(\mathrm{SW})^{s_1}} & \qw & \ghost{(\mathrm{SW})^{s_1\cdot \bar{s}_2}}      & \qw & \qw  & \quad \cdots \quad && \ghost{(\mathrm{SW})^{(\prod_{i=1}^{k-1} s_1)\cdot \bar{s}_k}} & \qw & \qw
}}
\]
The equivalent circuit without $Swap$ will be:
\[
\scalebox{0.8}{
\Qcircuit @C=1em @R=.7em {
 &  \gate{G_1^{\bar{s}_1}} & \gate{G_2^{\bar{s}_1 + \bar{s}_2}} & \qw & & \cdots \quad \quad & \gate{G_k^{\sum_{i=1}^k \bar{s}_i}} &\multigate{1}{(\mathrm{SW})^{\bigoplus_{i=1}^k s_i}} & \qw\\
 &  \gate{G_1^{s_1}} & \gate{G_2^{s_1 \cdot  s_2}} & \qw &  \quad \quad \cdots \quad \quad & & \gate{G_k^{\prod_{i=1}^k s_i}} & \ghost{(\mathrm{SW})^{\bigoplus_{i=1}^k a_i}} &\qw
}}
\]

This lets us convert the circuit in Fig. \ref{fig:ubqc_resource_set}(b) to the equivalent circuit without the $Swap$ gate as shown in Fig. \ref{fig:equiv_rz_gate}. 
The circuit without $Swap$ gates is just the expanded form of $R_z((-1)^{q_m}s_m \pi/2^m)$ as trivially visible from Theorem \ref{thrm1:rz_gate}.
Note, the effect on the ancilla qubit is omitted from the figure for the sake of simplicity.

\bibliography{main_ref}

\end{document}